\documentclass{aa}  

\usepackage{graphicx}
\usepackage{txfonts}
\usepackage{amsmath}
\usepackage{float}
\usepackage{subfig}
\usepackage{color}

\usepackage{xspace}

\newcommand{\teff}{$T_{\rm eff}$\xspace}

\newcommand{\logg}{$\log g$\xspace}

\newcommand{\vsini}{v $\sin i$\xspace}

\newcommand{\vmic}{v$_{mic} $\xspace}
\newcommand{\vmac}{v$_{mac}$\xspace}
\newcommand{\vrad}{v$_{rad}$\xspace}
\newcommand{\ha}{H$\alpha$\xspace}
\newcommand{\hb}{H$\beta$\xspace}
\newcommand{\kms}{km s$^{-1}$\xspace}

\begin{document}

   \title{Line-dependent veiling in very active classical T Tauri stars}

   %\subtitle{I. }

   \author{A.C.S. Rei\inst{1,2}
          \and
          P. P. Petrov\inst{3}
          \and
          J.F. Gameiro\inst{1,2}
          }

   \institute{Instituto de Astrof\'isica e Ci\^encias do Espa\c co, CAUP, Rua das Estrelas, PT4150-762 Porto, Portugal\\
         \and
             Departamento de F\'isica e Astronomia, Faculdade de Ci\^encias, Universidade do Porto, Rua do Campo Alegre, 4150-762 Porto, Portugal
         \and
             Crimean Astrophysical Observatory of Russian Academy of Sciences, p/o Nauchny, 298409, Republic of Crimea\\
             }

   \date{Accepted: 2017 November 22}

% \abstract{}{}{}{}{} 
% 5 {} token are mandatory
\abstract
{The  T Tauri stars with active accretion disks show veiled photospheric spectra. This is supposedly due to non-photospheric continuum
radiated by hot spots beneath the accretion shocks at stellar surface and/or chromospheric emission lines radiated by the post-shocked
gas. The amount of veiling is often considered as a measure of the mass-accretion rate.}
{We analysed high-resolution photospheric spectra of accreting T Tauri stars Lk\ha 321, V1331 Cyg, and AS 353 A  with the aim of clarifying the nature of the line-dependent veiling. Each of these objects shows a strong emission line spectrum and powerful wind features indicating high rates of accretion and mass loss.}
{Equivalent widths of hundreds of weak photospheric lines were measured in the observed spectra of high quality and compared with those in synthetic spectra of appropriate models of stellar atmospheres.  }
{The photospheric spectra of the three T Tauri stars are highly veiled. We found that the veiling is strongly line-dependent: larger in stronger photospheric lines and weak or absent in the weakest ones. No dependence of veiling on excitation potential within 0 to 5 eV was found. Different physical processes responsible for these unusual veiling effects are discussed in the framework of the magnetospheric accretion model.}
{ The observed veiling has two origins: 1) an abnormal structure of stellar atmosphere heated up by the accreting matter, and  2) a non-photospheric continuum radiated by a hot spot with temperature lower than 10000 K. The true level of the veiling continuum can be derived by measuring  the weakest photospheric lines with equivalent widths down to $\approx$10 m\AA. A limited spectral resolution and/or low signal-to-noise ratio results in overestimation of the veiling continuum. In the three very active stars, the veiling continuum is a minor contributor to the observed veiling, while the major contribution comes from the line-dependent veiling. }

   \maketitle
%
%________________________________________________________________

\section{Introduction}

%%%%%%%%%%%%%%%%%%%%%%%%%%%%%%%%%
\begin{figure*}
\centering
\includegraphics[width=0.96\textwidth]{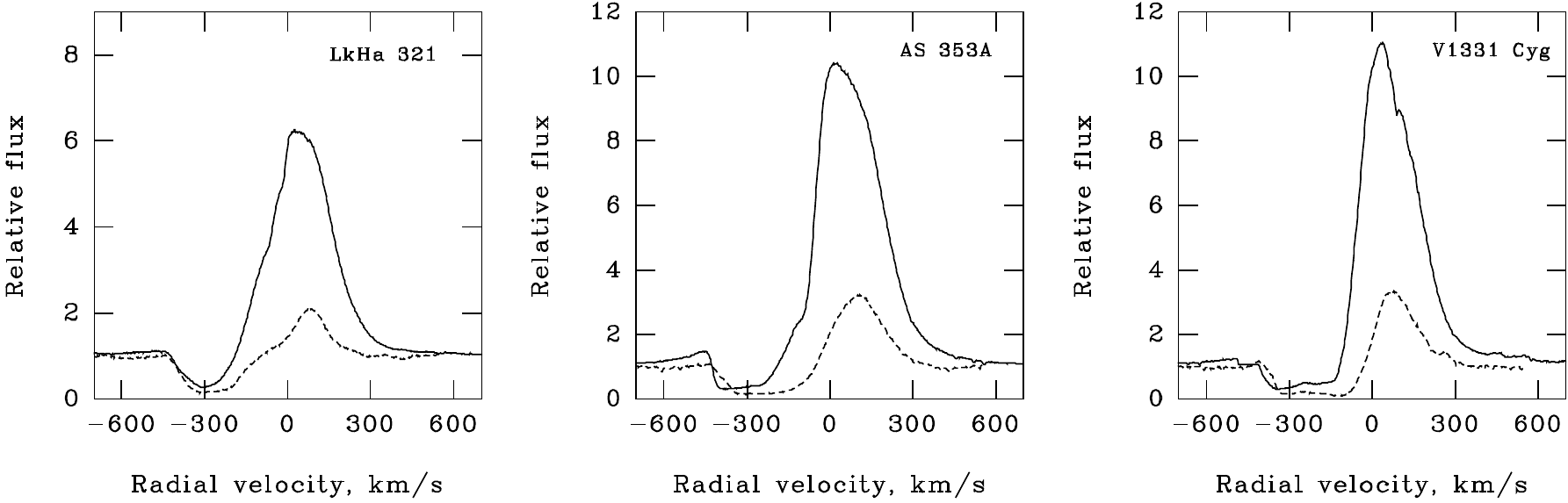}   
\caption{\ha (solid line) and \hb (dashed line) profiles.}                       
\label{fig1}
\end{figure*}
%%%%%%%%%%%%%%%%%%%%%%%%%%%%%

TT Tauri stars are late-type stars at the pre-main sequence (PMS) phase of evolution. These stars are in fact star-disk systems and their spectra show several features associated with the interaction between star and disk. There are two classes of T Tauri stars, based on the degree of activity of the star-disk system: classical T Tauri stars (cTTS) and weak-line T Tauri stars (wTTS). The rich emission spectrum of cTTS is related to magnetospheric accretion of matter from the inner disk regions onto the star. Yet in the early studies of cTTS it was found that the strength of the  photospheric lines is reduced as compared to a main-sequence (MS) star of the corresponding spectral type. \citet{joy49} noted that this may be due to emission within the lines and due to veiling of the spectrum by an overlying continuous radiation. 

In the concept of magnetospheric accretion, the veiling continuum is attributed to a shock-heated gas (a hot spot) at the stellar surface below the accretion stream. The veiling was often considered as a measure of accretion rate \citep[e.g.][]{basri90, hartigan91, hartigan95, valenti93, calvet04} along with other indicators, like the luminosity in \ha and other emission lines \citep[e.g.][]{muzerolle98, herczeg08, dahm08, fang09, rigliaco12, mendigutia15}. 

As more data were accumulated, some doubts appeared about the interpretation of the veiling origin. The expected correlation between the veiling in the optical spectrum and the brightness of the star was not found in some cTTS.  For example, in RW Aur A the veiling is highly variable from night to night. However, no correlation has been found with the stellar brightness in the V band from the photometric and spectroscopic monitoring of the star \citep{petrov01}. No rotational modulation of the veiling was reported for the star.

\citet{gahm08} showed that in RU Lup and S CrA SE large veiling factors correlate only weakly with brightness. In the case of S CrA SE a narrow emission core appears at the bottom of photospheric absorptions at moments of enhanced accretion. Furthermore, when the veiling is measured carefully in every single spectral line within a narrow wavelength range, it turns out that the veiling depends on the line strength in a template of the corresponding spectral type.  Stronger lines can be veiled considerably, while the weakest lines remain about normal for the spectral type. So far, line-dependent veiling has been found in DR Tau \citep{petrov11} and V1331 Cyg \citep{petrov14}. A similar effect was noticed earlier in DI Cep \citep{gameiro06}. Line-dependent veiling was found in spectra of EX Lup in quiescent state, when the accretion rate was relatively low \citep{sicilia15}. The authors concluded that photospheric lines in EX Lup are veiled by the broad emission lines from the extended pre-shock material. 

In order to clarify the nature of line-dependent veiling, we undertake detailed inspection of photospheric lines in cTTS with high accretion rates. In this work we investigate the veiling effect in spectra of three cTTS: Lk\ha 321 (=V1980 Cyg), AS 353 A (=V1352 Aql), and V1313 Cyg.  

%__________________________________________________________________

\section{Observational data}

In this research we use spectra obtained by George Herbig with the High Resolution Echelle Spectrometer (HIRES) at Keck-1\footnote{The W. M. Keck Observatory is operated as a scientific partnership between the California Institute of Technology, the University of California, and the National Aeronautics and Space Administration. The Observatory was made possible by the generous financial support of the W. M. Keck Foundation.} \citep{vogt94}. We estimated a spectral resolution of R$\sim$50000 near 6300 \AA\  from the weakest telluric lines in this region.

The spectra were obtained on 2002 December 16 (Lk\ha 321), 2003 July 06 (AS 353 A), and 2004 July 24 (V1331 Cyg). The wavelength coverage was 4350-6750 \AA~with some gaps between spectral orders in the region > 5000 \AA. One more spectrum of V1331 Cyg was taken on 2007 November 23 with the same spectrograph and a mosaic of three CCDs covering the spectral range 4750-8690 \AA. 
In the spectra of LK\ha 321 and V1331 Cyg, the signal-to-noise ratio per resolution element (S/N) is 170 at 5000\,\AA\, and rises to 400 at 6500\,\AA. The spectrum of AS353A has a S/N of 200 at 5000\,\AA\, and  350 at 6500\,\AA.

The spectra analysed belong to three cTTS with high accretion rates, whose stellar parameters are difficult to estimate. A summary of previous and current estimations of some stellar parameters for these stars can be found in Table \ref{table1}. The estimations of spectral type, effective temperature (\teff), projected rotational velocity (\vsini), radial velocity (\vrad), and equivalent widths of the \ha (EW(\ha)) and \ion{He}{I} 5875 \AA\ (EW(HeI)) lines for each star are described in Section 3.

Utrecht Echelle Spectrograph (UES) spectra of each star were also available, from observations done in November of 1998. The UES spectra have a similar resolution to those of HIRES, but with a S/N < 100. Although the S/N is not high enough to perform the same type of study done with the HIRES spectra, it is possible to check for a change in radial velocity of the stars from  spectra taken in different epochs.

\begin{table*}
\caption{Stellar parameters} % title of Table
\label{table1} % is used to refer this table in the text
\centering % used for centering table
\begin{tabular}{|c|ccccccccccc|} % centered columns (2 columns)
\hline\hline % inserts double horizontal lines
Star & Sp.Type & \teff  & log L$_{*}$ & M$_{*}^d$ & R$_{*}^d$ & \logg$^d$& \vsini & \vrad
(HIRES) & EW(\ha) & EW(HeI) \\ % table heading
  &  & (K) & (L$_{\odot}$) & (M$_{\odot}$) & (R$_{\odot}$) & (dex)&(\kms) & (\kms)& (\AA)
& (\AA)\\
\hline % inserts single horizontal line

Lk\ha 321 &    G5-G7 &  5500-5250  & 1.42$^a$ &   2.9    & 5.3 & 3.5 &   32 &
-16.0$\pm$0.5  & 30&  0.08 \\
V1331 Cyg & G7-K1  & 5250-5000   & 1.32$^b$  & 2.9   &  5.5 & 3.5 &  <6   &   
-15.0$\pm$0.3 & 53  & 0.22 \\
AS 353A   &   K0-K1  & 5100-4900   & 0.4$^c$ &  1.6  &  2.1 & 4.0 &  <6   & -10.4$\pm$0.2
  & 58  & 1.30 \\

\hline %inserts single line
\end{tabular}
\tablefoot{Sp.type, \teff, \vsini, \vrad, EW(\ha), and EW(HeI) are from this paper.
$^a$~\citet{cohen79}, $^b$~\citet{tokunaga04}, $^c$~\citet{hamann92}, $^d$~Obtained from 
\teff and L$_{*}$ and using the PMS models by \citet{siess00}.
}
\end{table*}

The three objects show strong emission line spectra, powerful wind features, and highly veiled photospheric spectra. Typically, in spectra of cTTS the emission lines of neutral and ionized metals exhibit a composite profile: a broad component formed in the accretion funnel and a narrow component consistent with an origin in the post-shocked gas near the stellar surface (e.g. \citet{beristain98}). This is the case of Lk\ha 321 and AS 353A, where the broad component is dominant with full width at half maximum (FWHM) $\approx$ 140 \kms. On the contrary, in V1331 Cyg the numerous emission lines of metals show only a single moderately broad profile with FWHM = 40-60 \kms (see Sect. 7.1.). 

The most prominent lines are the hydrogen Balmer and Paschen lines, the \ion{Ca}{II}, \ion{Na}{I} D, and \ion{O}{I} 8446 \AA~lines with FWHM $\approx$ 200 \kms. These broad lines are supposedly formed in the high velocity infalling gas at the base of the accretion column, before the shock (pre-shocked gas), within the stellar magnetosphere. The P Cyg type of the profiles of \ha and \hb lines indicates an extensive mass-loss (Figure \ref{fig1}).

\begin{figure*}
\centering
\includegraphics[width=0.95\textwidth]{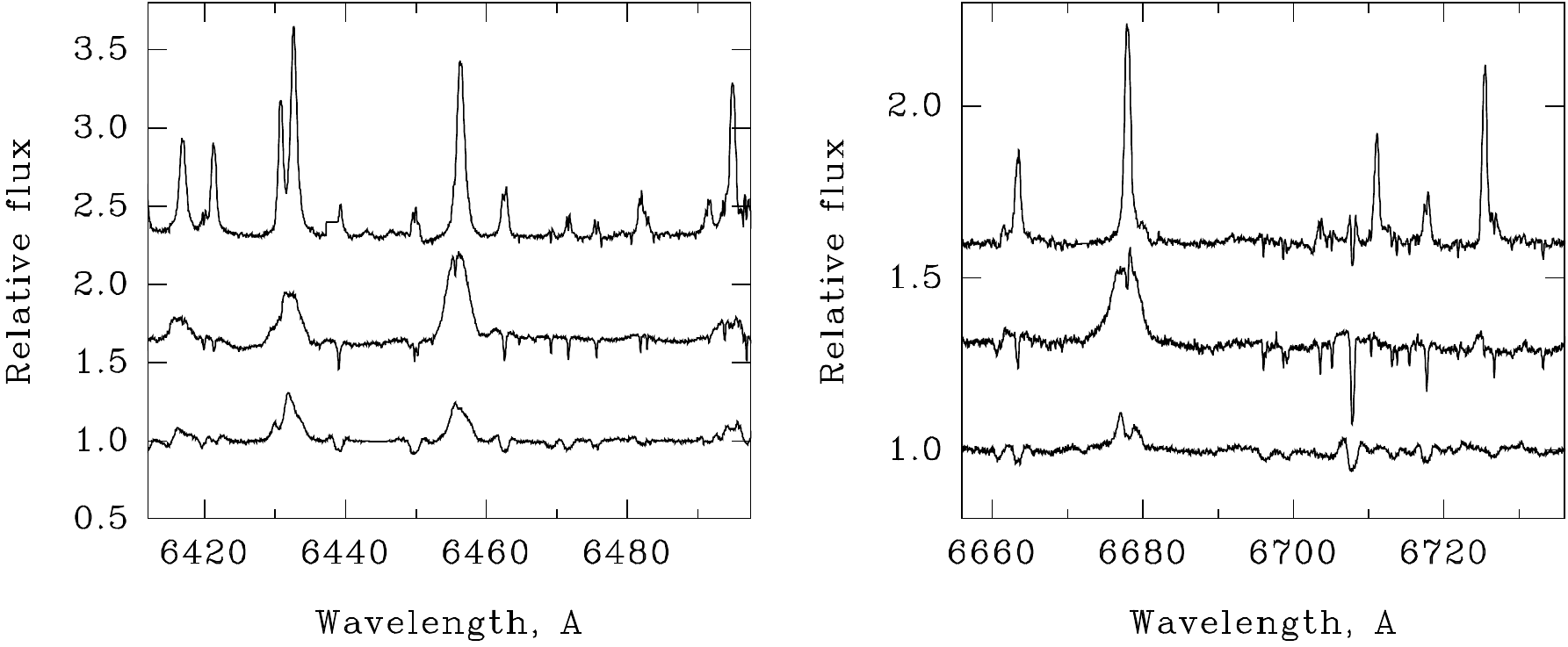}
\caption{Fragments of spectra: Lk\ha 321 (lower), AS 353A (middle), and V1331 Cyg (upper).}
\label{fig2}
\end{figure*}

Figure \ref{fig2} compares two spectral fragments of the  stars to show the differences in strength and width between the photospheric and emission lines. In all three objects the narrow forbidden emissions of [\ion{O}{I}] 6300.3 \AA~and 6363.7 \AA~are present.

%__________________________________________________________________

\section{Data analysis}

%%%%%%%%%%%%%%%%%%%%%%%%%%%%%%%%%%%%%%%
%%%%%%%%%%%%%%%%%%%%%%%%%%%%%%%%%%%%%

Template spectra of non-active main-sequence (MS) stars, wTTS, or synthetic spectra are needed to perform the analysis of cTTS.
In this investigation we analyse rather weak spectral lines, with equivalent width (EW) of about 10 m\AA, so we need high-resolution, high S/N spectra of template stars of different spectral type. In order to minimize the errors of measurements, we prefer to use a grid of synthetic spectra. This way we avoid possible spectral peculiarities from the real stars that could affect our results.

\subsection{Template synthetic spectra}

The synthetic spectra were produced by the software package Spectroscopy Made Easy (hereafter SME) \citep{valenti96}. 
It was initially used to confirm the spectral type of each cTTS, with the estimation of \teff, as well as for the determination of \vsini and \vrad.

Two sets of synthetic spectra with spectral resolution R = 50000 were produced. The first set corresponds to a grid of narrow-band spectra with several values of stellar parameters, which was used in the determination of spectral type, as well as \vsini and \vrad. Then, specific spectra for each star in a wider wavelength band between 4500 and 7000 \AA~were produced. This second set of synthetic spectra corresponds to our templates and was used to measure EWs of hundreds of spectral lines for comparison with the same lines in the spectra of cTTS.

The spectral type determination, described in detail in Section 3.2, allowed us to gather information mainly about \teff. The usual criteria of surface gravity (\logg) can hardly be used in the case of highly veiled spectra with strong lines in emission. Due to that we are assuming the value \logg = 3.75 dex for these stars (see Section 3.2). We also assume a priori a solar-like metallicity ([M/H] = 0.0).

Besides the mentioned stellar parameters, SME also needs input information about the microturbulence velocity (\vmic) and macroturbulence velocity (\vmac), as well as the instrumental resolution. For the \vmic parameter we assume a value that is typical for cool stars and wTTS \citep[e.g.][]{padgett96, rojas08, james06, taguchi09}: \vmic = 1.0 \kms. For the \vmac value we assume a value based on the \teff value, as reported by \citet{valenti05}. The three stars have spectral types between K1 and G5 corresponding to temperatures between 4900 K and 5500 K for sub-giant stars (Table \ref{table1}). For this range of temperatures we assume a value of \vmac = 2.8 \kms, consonant to an average \teff = 5000 K.

To create the synthetic spectra we use the Kurucz grids of stellar atmospheric models, considering local thermodynamic equilibrium (LTE) conditions. 
The line list used by SME to create the synthetic spectra was extracted from the Vienna Atomic Line Database (VALD) database using the “Extract stellar” option \citep{ryabchikova15}. Since we measured several hundred spectral lines, we did not apply any corrections to the log gf and Van der Waals damping parameters.

\subsection{Spectral type determination}

\begin{table*}
\caption{Photospheric line pairs used in the spectral type determination} % title of Table
\label{table:linepairs} % is used to refer this table in the text
\centering % used for centering table
\begin{tabular}{|c|ccc|ccc|} % centered columns (2 columns)
\hline\hline % inserts double horizontal lines
%\multicolumn{7}{c}{ Lk\ha 321}\\
%\hline
 Star& & Line 1 &  & & Line 2 & \\
&$\lambda$ (\AA) & Ele. & Exc. Pot. (eV)  & $\lambda$ (\AA) & Ele. & Exc. Pot. (eV) \\ % table heading 
\hline % inserts single horizontal line
&6013.416       &       \ion{Ti}{I}     &       1.07    &       6016.605        &       \ion{Fe}{I}     &       3.55    \\
&6039.729       &       \ion{V}{I}      &       1.06    &       6041.950        &       \ion{Fe}{I}     &       4.96    \\
Lk\ha 321&6325.164      &       \ion{Ti}{I}     &       0.02    &       6327.599        &       \ion{Ni}{I}     &       1.68    \\
&6469.193       &       \ion{Fe}{I}     &       4.48    &       6471.662        &       \ion{Ca}{I}     &       2.53    \\

%\hline\hline % inserts double horizontal lines
%\multicolumn{7}{c}{V1331 Cyg} \\
%\hline
%& & Line 1 &  & & Line 2 &\\
%&$\lambda$ (\AA) & Ele. & Exc. Pot. (eV)  & $\lambda$ (\AA) & Ele. & Exc. Pot. (eV) \\ % table heading 
\hline
&5054.074       &       \ion{Ti}{I}     &       2.68    &       5054.642        &       \ion{Fe}{I}     &       3.64    \\
&5289.269       &       \ion{Ti}{I}     &       0.84    &       5289.817        &       \ion{Y}{II}     &       1.03    \\
&5295.312       &       \ion{Fe}{I}     &       4.42    &       5295.776        &       \ion{Ti}{I}     &       1.07    \\
&5376.599       &       \ion{Ti}{I}     &       0.00    &       5376.830        &       \ion{Fe}{I}     &       4.29    \\
V1331 Cyg&5384.630      &       \ion{Ti}{I}     &       0.83    &       5385.575        &       \ion{Fe}{I}     &       3.69    \\
&5385.133       &       \ion{V}{I}      &       2.61    &       5385.575        &       \ion{Fe}{I}     &       3.69    \\
&5440.509       &       \ion{Ti}{I}     &       1.43    &       5441.339        &       \ion{Fe}{I}     &       4.31    \\
&5460.499       &       \ion{Ti}{I}     &       0.05    &       5461.549        &       \ion{Fe}{I}     &       4.45    \\
&5465.773       &       \ion{Ti}{I}     &       1.07    &       5466.987        &       \ion{Fe}{I}     &       3.57    \\

%\hline\hline %inserts single line
%\multicolumn{7}{c}{ AS 353A} \\
%\hline
%& & Line 1 &  & & Line 2 & \\
%&$\lambda$ (\AA) & Ele. & Exc. Pot. (eV)  & $\lambda$ (\AA) & Ele. & Exc. Pot. (eV) \\ % table heading 
\hline
&6111.070       &       \ion{Ni}{I}     &       4.09    &       6111.651        &       \ion{V}{I}      &       1.04    \\
&6116.180       &       \ion{Ni}{I}     &       4.27    &       6116.990        &       \ion{Co}{I}     &       1.79    \\
&6116.180       &       \ion{Ni}{I}     &       4.27    &       6119.528        &       \ion{V}{I}      &       1.06    \\
AS 353A&6146.207        &       \ion{Ti}{I}     &       1.87    &       6147.834        &       \ion{Fe}{I}     &       4.08    \\
&6151.617       &       \ion{Fe}{I}     &       2.18    &       6152.292        &       \ion{Si}{I}     &       5.96    \\
&6154.225       &       \ion{Na}{I}     &       2.10    &       6155.134        &       \ion{Si}{I}     &       5.62    \\
&6156.023       &       \ion{Ca}{I}     &       2.52    &       6157.727        &       \ion{Fe}{I}     &       4.08    \\

\hline\hline %inserts single line
\end{tabular}
\tablefoot{Some of the lines listed in this table are blends, particularly for the Lk\ha 321 spectrum.}
\end{table*}

\begin{table*}
\caption{Template synthetic spectra parameters for each star } % title of Table
\label{table2} % is used to refer this table in the text
\centering % used for centering table
\begin{tabular}{|c|cccccc|} % centered columns (2 columns)
\hline\hline % inserts double horizontal lines
Template & \teff & \logg  & [M/H] & \vsini & \vmic & \vmac \\ % table heading 
 for  &  (K) & (dex) & (dex) & (\kms)  & (\kms) & (\kms)  \\
\hline % inserts single horizontal line
Lk\ha 321 &  5250  & 3.75   &  0.0   & 32   &  1.0 & 2.8   \\
V1331 Cyg  & 5000   & 3.75  &  0.0  & 1  & 1.0  & 2.8  \\
AS 353A    & 5000   & 3.75     & 0.0  &  1   & 1.0 & 2.8  \\
\hline %inserts single line
\end{tabular}
\end{table*}

The spectral types of our targets were determined by using a grid of synthetic spectra with several values of \teff and \logg.
Determination of the spectral type of cTTS is hampered by the veiling effect. As will be shown below, the stronger lines are most affected by the veiling, therefore the use of strong lines in the spectral classification should be avoided. To perform this task we identified the less blended photospheric lines on several spectral fragments throughout the whole spectrum.

To find the value of \teff for each star, we identified several pairs of weak  lines with ratios sensitive to temperature changes in the synthetic spectra grid. The synthetic spectra with the ratios of lines similar to those of the cTTS were chosen as the appropriate templates. Typically, the cTTS spectra drop in between two synthetic templates with 250 K difference. The pairs of photospheric lines used to determine the spectral type of each cTTS are listed in Table \ref{table:linepairs}.

Due to the numerous emission lines and the veiling of photospheric lines, it is problematic to derive the \logg parameter directly from the spectra of very active cTTS. It can be estimated from the grid of models of PMS stars \citep{siess00} with input parameters \teff and  L$_{*}$  (see Table \ref{table1}). For the synthetic templates we adopted log g = 3.75 dex. A simple test with the templates showed that differences of 0.25 dex in this parameter are not critical for our analysis of equivalent widths.

For the determination of the \vsini value, artificial veiling must be added to the synthetic spectra in order to mimic the real spectra. Because the estimation of veiling also depends on the \vsini value, this process must be done in an iterative way, until the synthetic spectra lines have a good match with the real ones.

The information regarding the synthetic templates for each cTTS, based on the previous estimation of stellar parameters, is summarized in Table \ref{table2}. Details on the individual spectrum of each star can be found in Sections 4, 5, and 6.

\subsection{Equivalent widths measurements}

Our study is based on the comparison of EWs of several hundred photospheric lines of cTTS with those of synthetic templates. To measure the EWs we used IRAF.\footnote{IRAF is distributed by the National Optical Astronomy Observatories, which are operated by the Association of Universities for Research in Astronomy, Inc., under cooperative agreement with the National Science Foundation.} Due to the presence of strong veiling and emission in these stars' spectra, only lines that could be clearly identified with their counterparts in the synthetic templates were measured. Regions with broad emission were avoided. In total, we measured the EWs of more than 500 photospheric lines, except for Lk\ha 321, where many lines are blended because of the large \vsini. We measured EWs of the less blended lines in Lk\ha 321 and in the corresponding template with the same \vsini.

Although the precision of the measurements is limited by S/N, the main source of errors is the uncertainty in the local continuum level, caused by the line blends of both photospheric absorptions and broad emissions. We use the synthetic template to find the adjacent points of the continuum for each photospheric line, and select the spectral intervals where the broad emissions are relatively small, not exceeding a few percent above the continuum level. The relative error in the EW of photospheric lines was found to be within 5-12 $\%$.

\subsection{Equivalent widths ratio as a measure of veiling} 

Traditionally, on cTTS the veiling factor (VF) is expressed as VF = EW(template)/EW(tts) - 1. In this expression it is assumed that EW(tts) is reduced by an additional (non-photospheric) continuum, and the VF is a measure of this non-photospheric continuum in units of the photospheric continuum. Since we do not know a priori the nature of the veiling, we will use in the following analysis only the ratio of EWs: EW(template)/EW(tts). In the veiled spectrum of cTTS, for most of the photospheric lines this ratio is larger than unity, although for the weakest lines it may be less than unity because of the measurement errors. The error in spectral type may also result in some bias. In the optical spectrum of cTTS, the veiling usually rises towards the blue region, which indicates the presence of a hot continuum \citep{basri90, calvet98}. In this study we also analyse how the veiling changes with wavelength. 

We used real spectra of the Sun and a G7IV star (HD190248) to evaluate how large the scatter of the EW(synthetic)/EW(real) is along the spectrum. The Sun spectrum\footnote{Source: http://www.blancocuaresma.com/s/benchmarkstars.} was obtained by High Accuracy Radial velocity Planet Searcher (HARPS) and has a S/N > 500 \citep{blanco14}. The G7IV star spectrum was extracted from the ESO Archive. We assigned a synthetic template with \teff = 5000 K and \logg = 3.75 dex to the G7IV star and a template with \teff = 5750 K and \logg = 4.4 dex to the Sun, both with solar metallicity. The EWs of the same lines were measured on both real and synthetic spectra and the ratio EW(synthetic)/EW(real) was analysed.

We found that the EW(synthetic)/EW(real) ratio does not depend on wavelength, line strength, and excitation potential of low level of the transition of the transition. Other atomic parameters, like excitation potential of the upper level and total excitation energy, have also been tested and no dependence with EW(synthetic)/EW(real) ratio was found. The line-to-line scatter of this ratio can be quantified with a standard deviation from the average. We found that the sample standard deviation is about 0.2 - 0.3 dex for the weakest lines with EW < 50 m\AA\ and 0.1 - 0.2 dex for stronger lines. There are two possible causes for the dispersion of values: one is due to the error in measurement and determination of continuum level, and another is due to the lack of correction for the log gf and Van der Waals damping parameters in the synthetic templates. Although the line-to-line scatter of the EW(synthetic)/EW(real) ratio is relatively large, it does not affect our results due to the large number of lines used in the analysis. 

%__________________________________________________________________

\section{Lk\ha 321}

%%%%%%%%%%%%%%%%%%%%%%%%%%%%%%%%%%%%%
\begin{figure}
\centering
\includegraphics[width=0.49\textwidth]{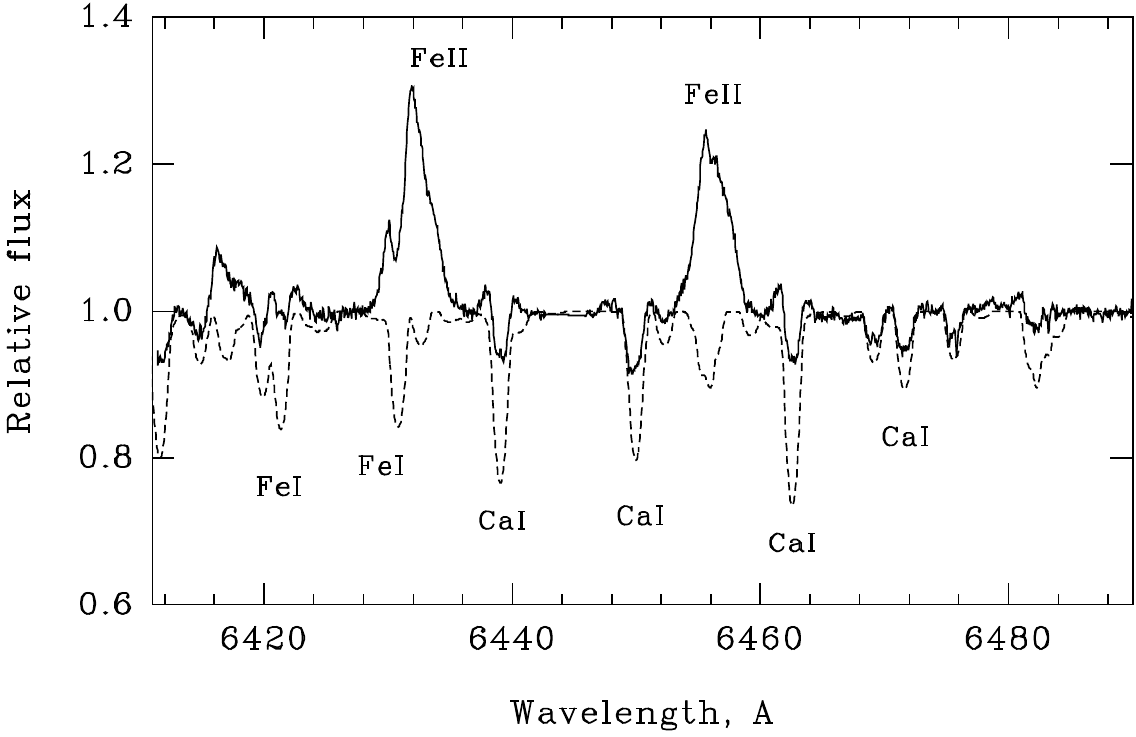}
\caption{Fragment of spectra of Lk\ha 321 (solid line) and the synthetic template \teff = 5250 K, \logg = 3.75 dex, \vsini = 32 \kms (dotted line). The two broad emissions belong to \ion{Fe}{II}. The weak broad emission wings can be noticed in the strongest \ion{Ca}{I} lines.}
\label{fig3}
\end{figure}
%%%%%%%%%%%%%%%%%%%%%%%%%%%%%%%%%%%%%

\begin{figure}
\centering
\includegraphics[width=0.49\textwidth]{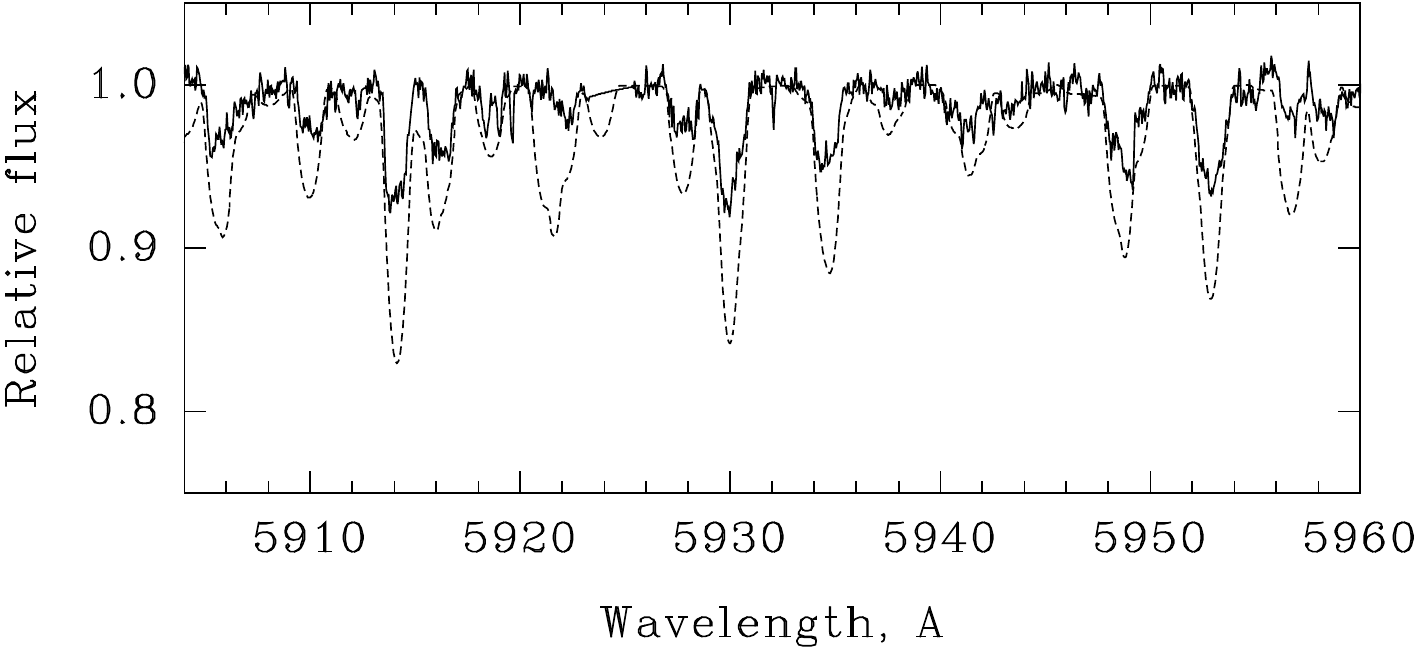}
\caption{Comparison of the photospheric lines in Lk\ha 321 and in the synthetic template. Most of the lines are of \ion{Fe}{I}. No broad emissions in this region, but the photospheric lines are veiled considerably.}
\label{fig4}
\end{figure}
%%%%%%%%%%%%%%%%%%%%%%%%%%%%%%%%%%%%%%%

\begin{figure}
\centering
\includegraphics[width=0.47\textwidth]{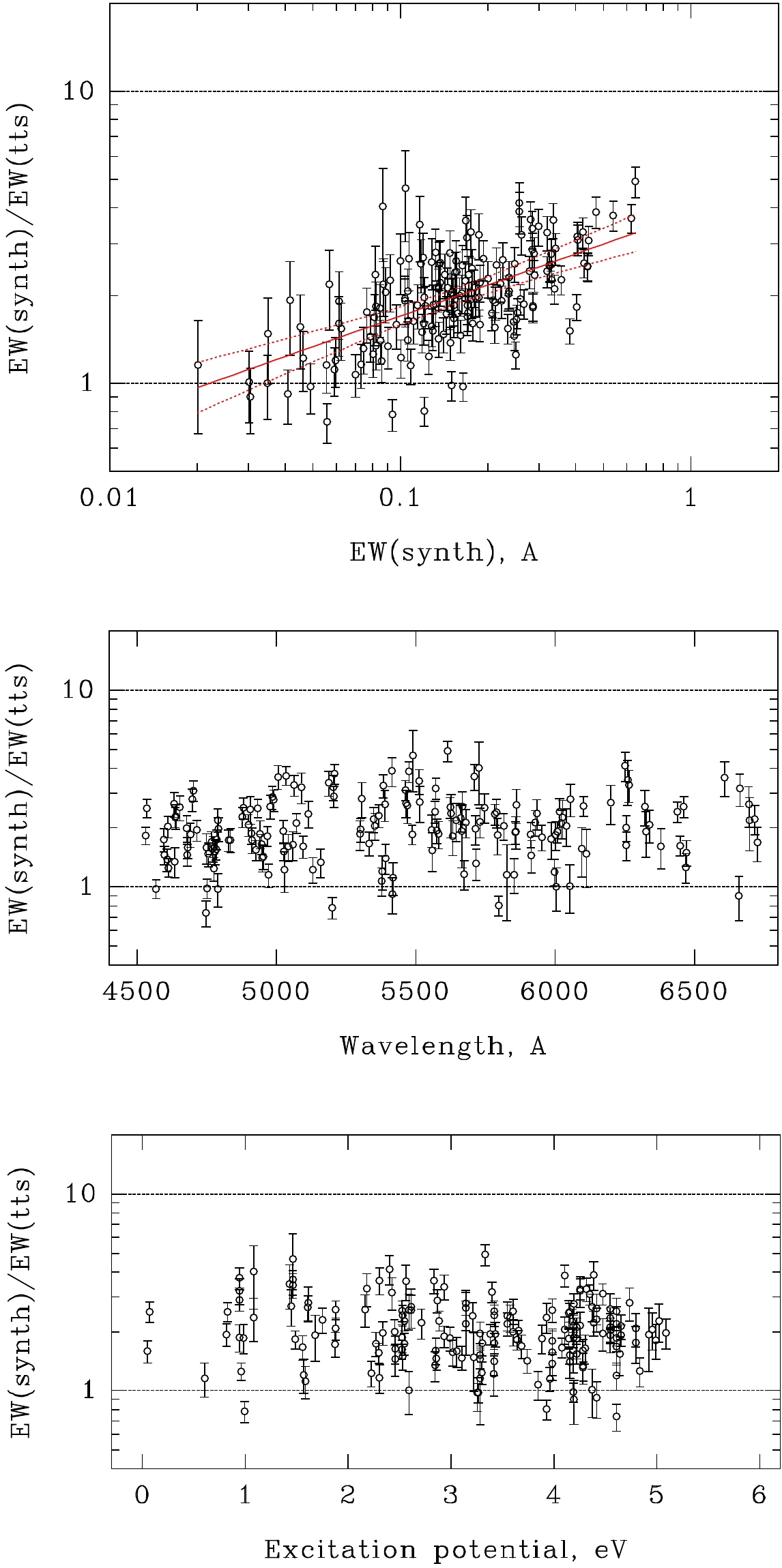}
\caption{Comparison of Lk\ha 321 with template \teff = 5250 K. Veiling  as a function of line strength (upper), wavelength (middle), and excitation potential (lower). The line of linear regression and the 99$\%$ confidence interval are indicated on the top panel. }
\label{fig5}
\end{figure}

Two of our objects, Lk\ha 321 and V1331 Cyg, are located in a dark cloud in Cygnus, a few degrees north from the star forming region NGC7000/IC5070. The distance to these two objects was previously estimated as 700 pc \citep{chavarria81} but later revised to 550 $\pm$ 50 pc \citep{shevchenko91}. 

The star Lk\ha 321 permanently shows strong wind features in the P Cyg profiles of the Balmer lines. The [\ion{S}{II}] images show a knotty jet extending 22 arcsec \citep{mundt98}. The forbidden line of [\ion{O}{I}] 6300 \AA\, has the central peak and the blue-shifted emission component at a radial velocity of about -376 \kms with respect to the star. 

This star shows low photometric variability, within visual magnitude V = 12.25 - 12.44 \citep{grankin07}. To date, no period of axial rotation has been reported. According to \citet{cohen79}, this star has spectral type G1 and an interstellar extinction A$_V$ = 2.23 $\pm$ 0.3 mag. The stellar bolometric luminosity estimated from these photometric data, assuming a distance d = 550 $\pm$ 50 pc, is: $\log(L*/L_{\odot})$ = 1.42 $\pm$ 0.12.

Figures \ref{fig3} and \ref{fig4} show fragments of the spectrum of Lk\ha 321, including emission and photospheric absorption lines. The emission component of neutral metals is broader than the photospheric absorptions and  becomes more noticeable in relatively strong transitions, where the photospheric counterparts have  EW $\geq$ 100 m\AA.  In even stronger lines the broad emission is well above continuum, with the photospheric absorption on top of it. The emission lines of ions have no photospheric counterpart and display a triangular profile.

The comparison of the Lk\ha 321 spectrum with the grid of synthetic templates results in the best fit to \teff = 5250 K and \vsini = 32 \kms as reported  in Table \ref{table2}. No significant change was observed between the radial velocities measured in UES spectra, \vrad = -14.4$\pm$1.0 \kms and in the  HIRES spectra (Table \ref{table1}).

After the construction of the synthetic template we were able to measure EWs of photospheric lines on both Lk\ha 321 and template spectrum, as described in Section 3.3. We investigate the dependence of the ratio EW(template)/EW(tts) on the line strength, wavelength, and excitation potential of the lower level of the transition.

Figure \ref{fig5} shows that the amount of veiling in a photospheric line is dependent on the strength of the line in the template spectrum (corresponding to a spectral type G8). The dependence remains even if we adopt an earlier spectral type, G0. In this case the average level of veiling becomes lower. However, some lines in Lk\ha 321 are stronger than in a G0 template, which results in a false negative veiling and indicates that an inappropriate  template star has been chosen. On the other hand, the lower limit of the spectral type is K1, because the low-excitation (0-1 eV) lines of neutrals are absent in the observed spectrum. Therefore, the dependence of veiling on line strength, shown in Figure \ref{fig5}, is quite robust against any inappropriate spectral type determination. 

The main result is that the strongest transitions are most affected  by the veiling. We do not find any veiling dependence either on excitation potential or wavelength (Figure \ref{fig5}). In this spectrum, the broad emission lines are unlikely to cause the line-dependent veiling. The weak broad emission can be noticed starting from EW of about 100 m\AA\, in the template spectrum, while the veiling is already strong in these lines. There must be another cause of the veiling.
Although the photospheric line width is relatively large (\vsini = 32 \kms), we do not see narrow emission cores there, like those observed in S CrA SE \citep{gahm08} and modelled by \citet{dodin12}.

%__________________________________________________________________

\section{V1331 Cyg}

The unusually strong emission and wind features in V1331 Cyg make it similar to the pre-outburst spectrum of the FU Orionis type star (FUor) V1057 Cyg. For that reason, V1331 Cyg was considered as a possible pre-outburst FUor \citep{welin76}. The star is viewed pole-on with a collimated outflow (jet) towards the observer \citep{mundt84, mundt98}. Furthermore, the star is surrounded by ring-like nebulae, which might be made up of remnants of powerful mass-loss events in the past \citep{kuhi64, mundt98, quanz07, choudhary16}. 

The Keck spectra of V1331 Cyg, used in our research, was previously analysed by \citet{petrovbabina14} and \citet{petrov14}, where the presence of the line-dependent veiling was demonstrated. In addition, the spectral type of G7-K0IV, \logg $\sim$ 3.5 dex, radial velocity \vrad = -15.0 \kms and a small vsini < 6 \kms were derived from the photospheric lines.
From the analysis of the forbidden line profiles and the blue-shifted ''shell'' components of strong permitted lines, it was proved that the star is viewed through the jet \citep{petrov14}. The forbidden line of [\ion{O}{I}] 6300 \AA~ has a strong central peak and a blue-shifted one at -235 \kms.

The star varies in brightness within V = 11.7-12.5 \citep{kolotilov83, shevchenko91} and to the best of our knowledge no period related with rotation has been reported. \citet{hamann92} estimated a stellar bolometric luminosity of $L = 21 L_{\odot}$, with interstellar extinction A$_V$ = 1.4 mag. The two spectra of 2004 and 2007 show no difference in strength of photosphere lines. The radial velocity of the star  is the same in both spectra and slightly smaller in the UES spectra obtained in 1998, \vrad = -16.8$\pm$0.3 \kms.

Spectral type analysis for V1331 Cyg places this star between the synthetic spectra with \teff = 5250 K and 5000 K, making it a G7IV-K1IV star (Table \ref{table1}). This result is consistent with previous determinations by \citet{petrovbabina14}. The complete set of parameters for the V1331 Cyg synthetic template are listed in Table \ref{table2}.

\begin{figure}
\centering
\includegraphics[width=0.47\textwidth]{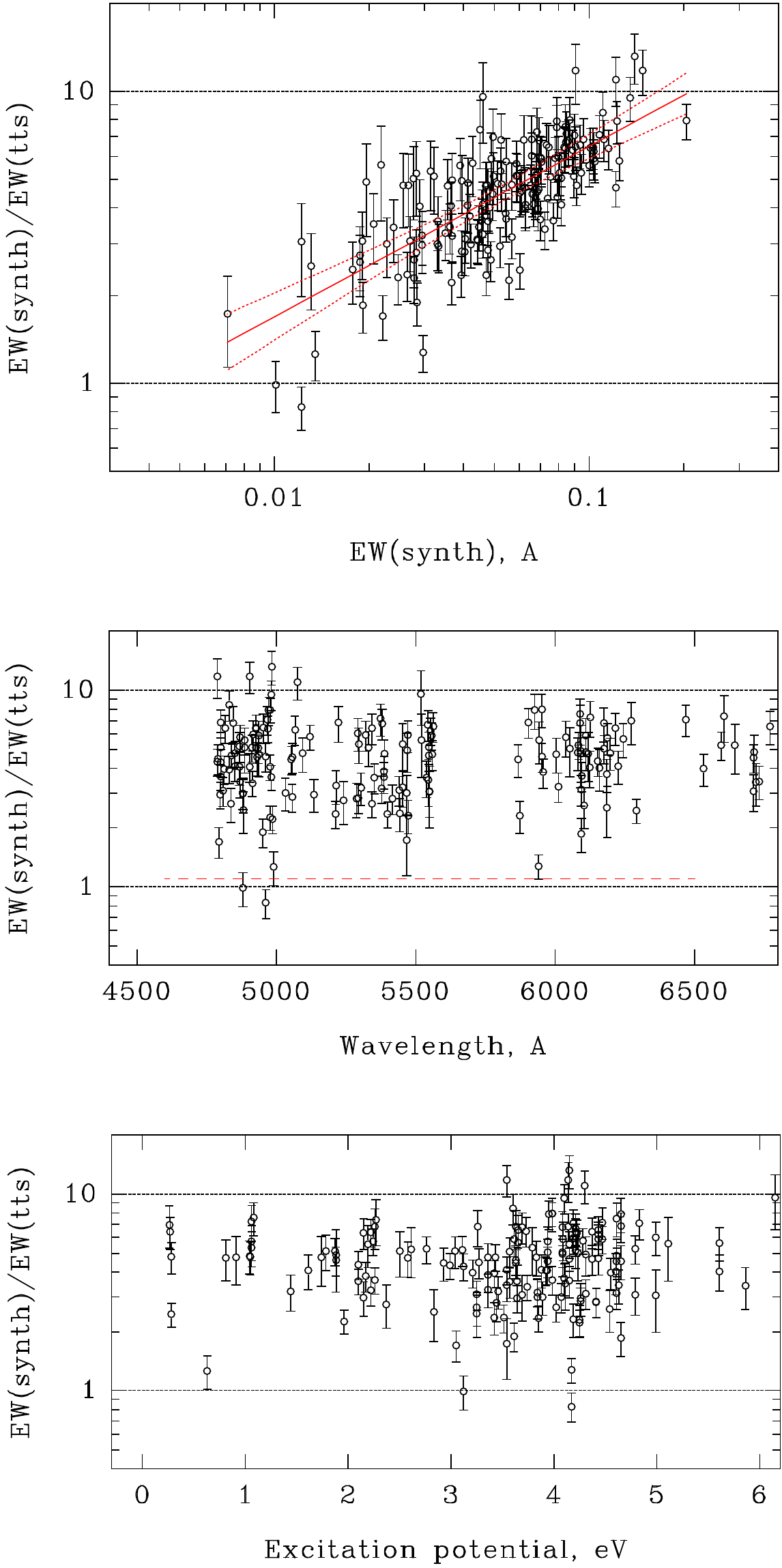}
\caption{Comparison of V1331 Cyg with template \teff = 5000 K. Veiling  as a function of line strength (upper), wavelength (middle), and excitation potential (lower). The line of linear regression and the 99$\%$ confidence interval are indicated on the top panel. Approximate level of the veiling continuum is indicated with the dashed line in the middle panel.}
\label{fig6}
\end{figure}

In Figure \ref{fig6} we present  the dependence of the ratio EW(template)/EW(tts) on the line strength, wavelength, and excitation potential of the lower level of the transition for V1331 Cyg. The top panel in Figure \ref{fig6} is similar to that obtained by \citet{petrov14}, where spectra of real stars were used as templates. Also, there is no dependence of the ratio EW(template)/EW(tts) on wavelength or excitation potential. The gaps along the wavelength scale on the middle panel of Figure \ref{fig6} reflect the regions of intense emission lines and gaps between spectral orders.

%%%%%%%%%%%%%%%%%%%%%%%%%%%%%%%%%%
%%%%%%%%%%%%%%%%%%%%%%%%%%%%%%%%%%%

%__________________________________________________________________

\section{AS 353A}

The star AS 353A belongs to the Aquila star forming region. The distance is estimated as d = 150 $\pm$ 50 pc \citep{prato03, rice06}. It is a mixed system with a primary cTTS, AS 353A, and a secondary wTTS, AS 353B. The secondary, located at 5.6" of the primary, is resolved into a subarcsecond binary, AS 353 Ba and Bb, separated by 0.24" \citep{tokunaga04}.  

The star AS 353A has a rich emission line spectrum with strong blue-shifted absorptions in all the Balmer lines, indicating a powerful outflow \citep{herbig83, eisloffel90, hamann92, alencar00}. The star drives the Herbig-Haro object HH-32 \citep{herbig83}. The forbidden line of [\ion{O}{I}] 6300 A has the central peak and two components at -270 and +250 \kms.

\citet{basri90} estimated the spectral type K2 from the optical spectroscopy. This was later confirmed from the K-band spectroscopy \citep{tokunaga04}. From the H-band spectroscopy, \citet{rice06} derived the following parameters for AS 353A: spectral type K5,  \vsini = 10 \kms , and stellar radial velocity \vrad = -11.4 $\pm$ 1.1 \kms. The stellar bolometric luminosity is $\log (L*/L_{\odot}) = 0.4$ \citep{tokunaga04}. Photometric variability within 0.86 mag in V was observed by \citet{fernandez96}. No periodicity related to stellar rotation was reported.

\begin{figure}
\centering
\includegraphics[width=0.49\textwidth]{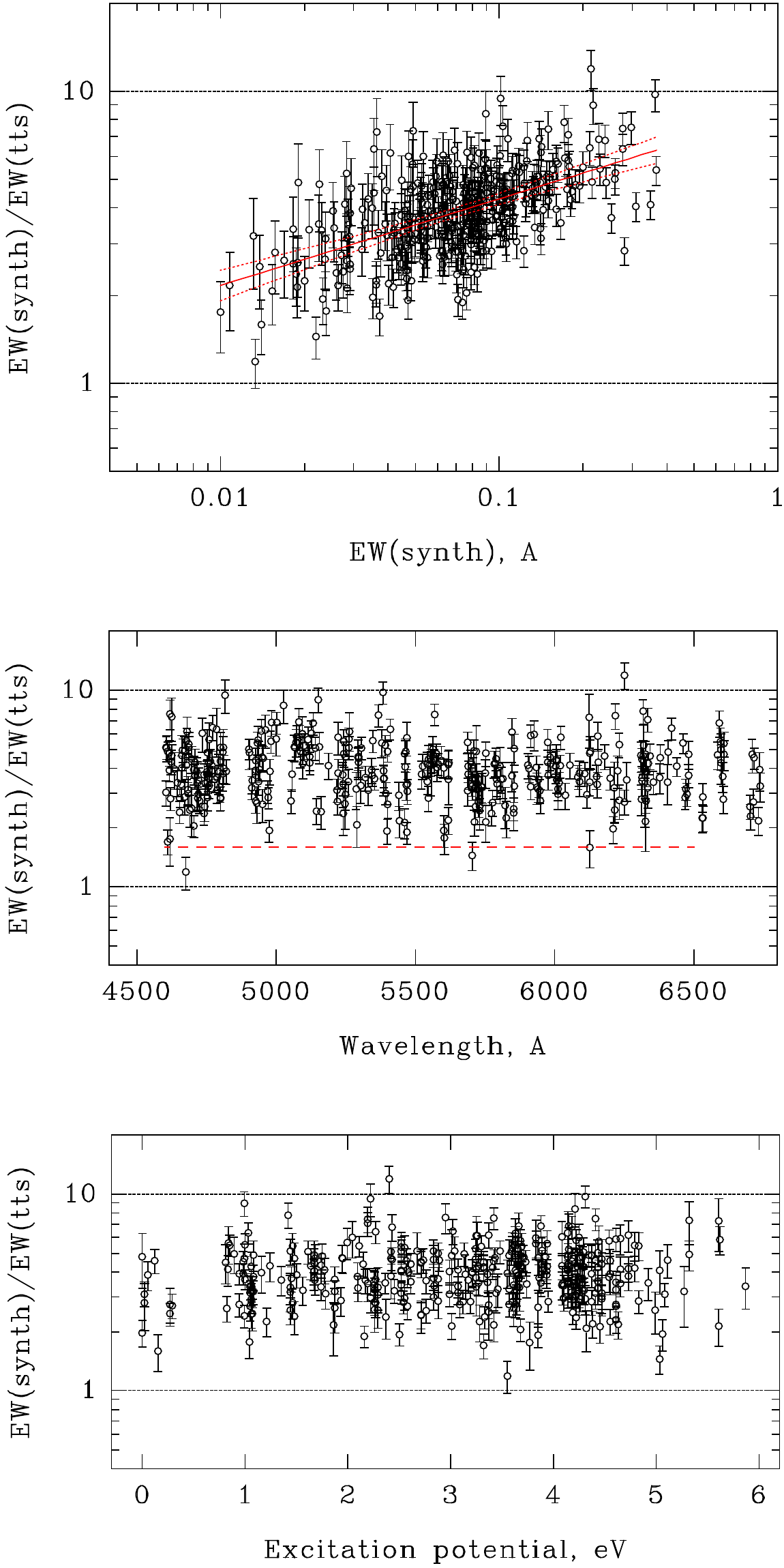}
\caption{Comparison of AS 353A with template \teff = 5000 K. Veiling  as a function of line strength (upper), wavelength (middle), and excitation potential (lower). The line of linear regression and the 99$\%$ confidence interval are indicated on the top panel. Approximate level of the veiling continuum is indicated with the dashed line in the middle panel.}
\label{fig7}
\end{figure}

Our spectral type analysis placed AS 353A between the synthetic templates with \teff = 4900 K and 5100 K, corresponding to a K0\,IV - K1\,IV star. The complete information about the AS 353A synthetic template can be found in Table \ref{table2}. From the HIRES spectra we measured similar radial velocity to that given by \citet{rice06} (see Table~\ref{table1}), but in the UES spectra we found higher velocity,  \vrad = -8.0$\pm$0.2 \kms.
This star also shows dependence of the ratio EW(template)/EW(tts) on the line strength (Fig. \ref{fig7}). Again, we found no correlation between the ratio EW(template)/EW(tts) with wavelength and excitation potential of the lower level of the transition.
The spectrum of AS 353A is remarkable in the combination of the narrow photospheric lines, \vsini < 6 \kms, and the broad emissions of metals, FWHM $\sim$ 150 \kms (Fig. \ref{fig9}). This makes clear the presence of the photospheric lines on top of the broad emissions.

%%%%%%%%%%%%%%%%%%%%%%%%%%%%%%%%%

\begin{figure}
\centering
\includegraphics[width=0.49\textwidth]{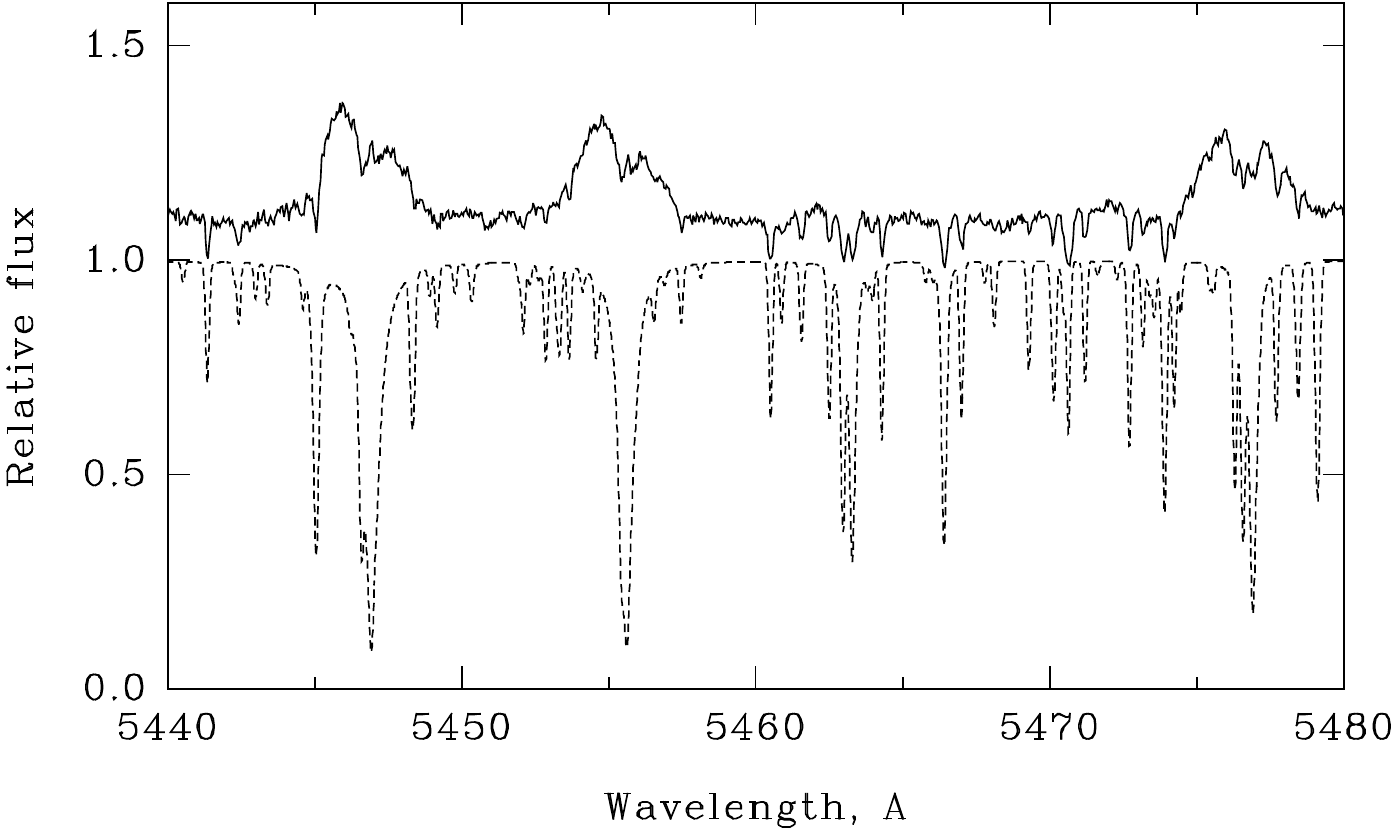}
\caption{Spectral fragments of AS 353A and the corresponding synthetic template (dashed line). We note the photospheric absorption lines on top of the broad emissions of \ion{Fe}{I}.}
\label{fig8}
\end{figure}

\begin{figure}
\centering
\includegraphics[width=0.49\textwidth]{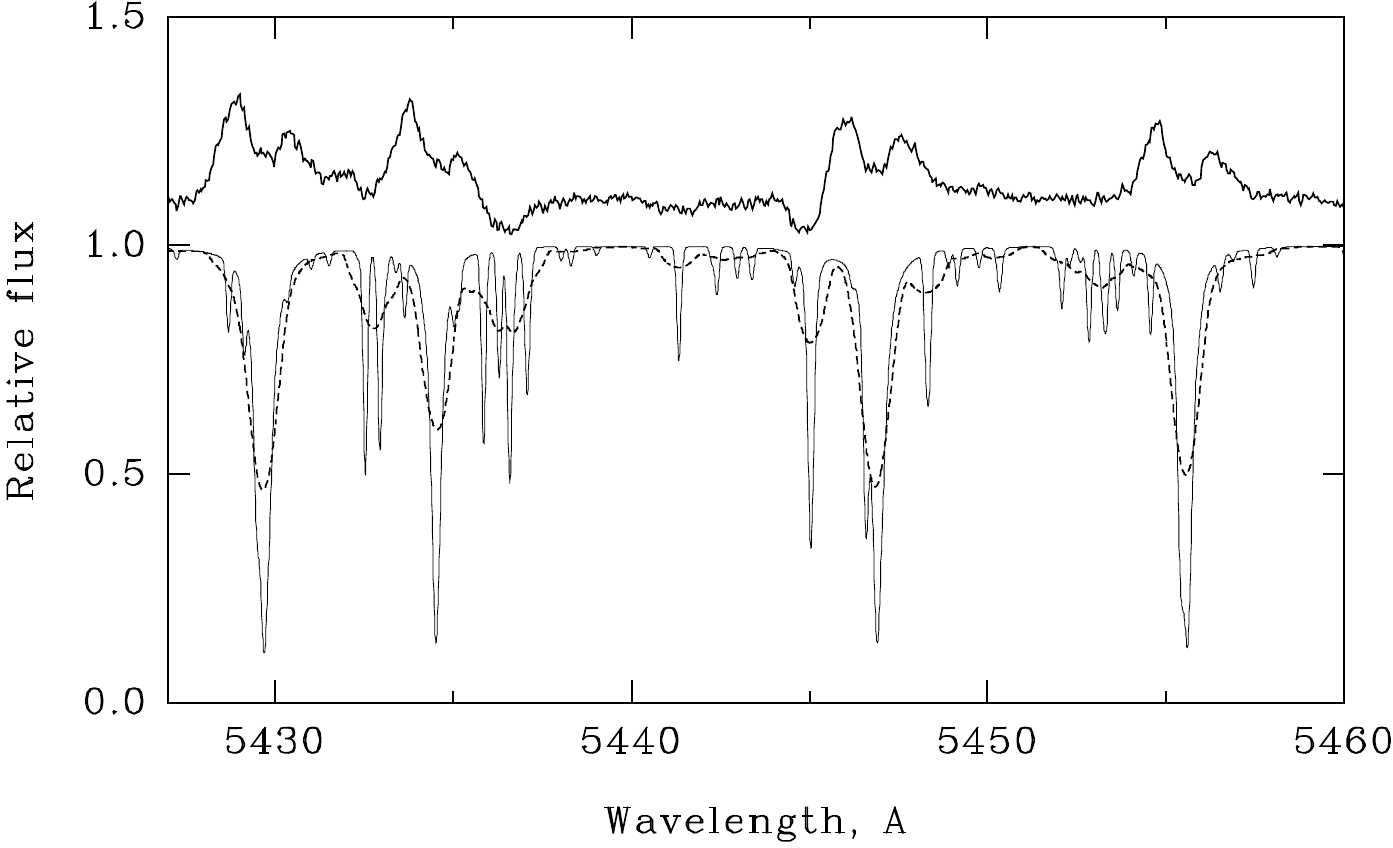}
\caption{Spectral fragments of Lk\ha 321 and the corresponding synthetic template (thin line). The template broadened to \vsini = 32 \kms is overplotted with dashed line. We note the photospheric absorption lines on top of the broad emissions of \ion{Fe}{I}.}
\label{fig9}
\end{figure}

\begin{figure}
\centering
\includegraphics[width=0.49\textwidth]{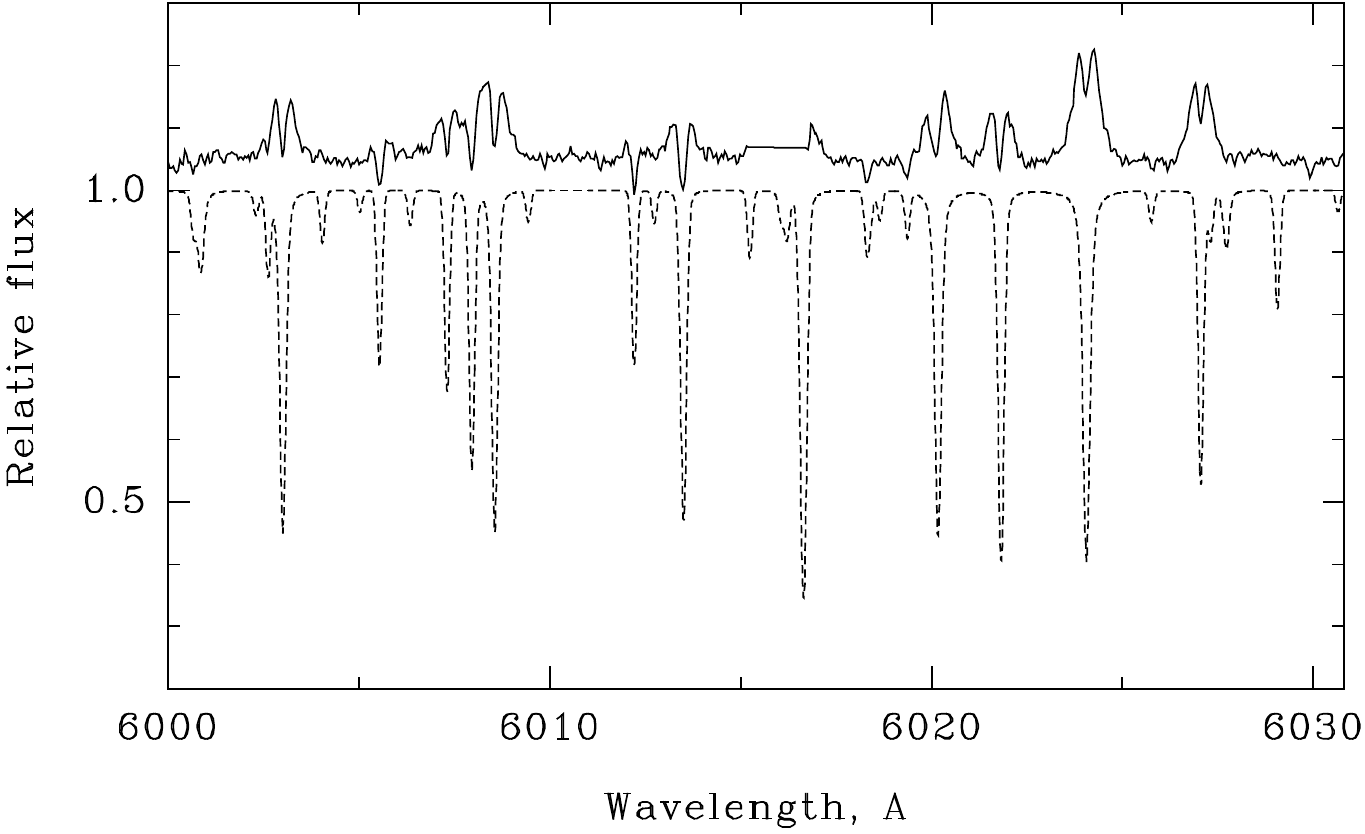}
\caption{Spectral fragment of V1331 Cyg and the corresponding synthetic template (dashed line). We note the photospheric absorption dips on top of the emission lines of \ion{Fe}{I} and \ion{Mn}{I}.}
\label{fig10}
\end{figure}
%%%%%%%%%%%%%%%%%%%%%%%%%%%%%%%%%%%%%%%%%%%%%%%%%%%

%__________________________________________________________________
 \section{Discussion}

Three possible sources of veiling were considered in the literature: 1) the veiling continuum radiated by a hot spot(s) at the base of accretion funnel(s) \citep[e.g.][]{basri90}, 2) narrow emission lines originating from the post-shocked gas \citep{dodin12}, and 3) broad emission lines originating in the infalling pre-shocked gas within the magnetosphere \citep{sicilia15}. 

In all of our objects the different spectral lines show different veiling, with a clear dependence of veiling on the strength of transition. This cannot be explained entirely with the traditional model of a hot spot radiating the veiling continuum. The three cTTS discussed here are outstanding  due to their high luminosity and high accretion rates, therefore the origin of the veiling  may not be necessarily the same as in a moderately accreting cTTS.
In the following we consider in more detail the possible sources of veiling.

\subsection{Broad emission}

Line-dependent veiling was investigated by \citet{sicilia15} from the spectral analysis of the EXor EX Lup in quiescence time. The authors suggested  that broad emission lines, originating from the pre-shocked material, can fill in the photospheric absorptions, thus resulting in line-dependent veiling. We examined whether this interpretation may be applied for our three stars.

In  AS 353A and Lk\ha 321, the broad emission lines of metals are much broader than the corresponding photospheric lines (see Figures \ref{fig8} and \ref{fig9}). The photospheric counterparts of the broad emissions of neutral metals (mostly \ion{Fe}{I})  are well seen on top of the emissions  and are  reduced in depth by about the same factor as the other strong photospheric lines  without the broad emission component.
It means that the broad emission component does not contribute considerably to the line-dependent veiling. The observed profile is just a sum of the optically thin broad emission and the underlying photospheric absorption. The only way the broad emission can affect the photospheric lines is that blends of the broad emissions may form an additional local pseudo continuum. In our analysis of the veiled photospheric lines, we avoided those lines which might be contaminated by such blends.

The third of our objects, V1331 Cyg, is remarkable for its pole-on orientation towards the observer. The low excitation emission lines of neutral and ionized metals appear at stellar velocity, and look similar in the three spectra taken in 2002, 2004, and 2007. These emission lines appear as not broadened. broadened by the flows within the stellar magnetosphere, probably because of the pole-on orientation of the star. 

\subsection{Photospheric hot spot}

Our results, presented above, show that the veiling continuum can be quantified only by the weak photospheric lines. Measuring the weakest lines, we probed the deepest layers of line formation. We also found that the veiling continuum determination is affected by the spectral resolution and S/N level. The spectra with lower resolution and/or poor S/N ratio lead to an overestimation of the veiling continuum. 

The hot spot hypothesis assumes that the accretion luminosity and the veiling continuum can be explained as being due to radiation of a hot spot, with temperature $T_{spot}$, and a filling factor, $f$,  representing  the relative size of hot spot region with respect to the stellar surface. The accretion luminosity can be estimated from the observed emission lines. It has been shown that $L_{acc}$ is well correlated with the line luminosity, $L_{line}$, of several emission lines, though the uncertainties in the empirical relations are in some cases high  (e.g \cite{fang09,rigliaco12,alcala14} and references therein). In our study we opted for the  empirical relations $L_{acc} - L_{line}$ obtained by \citet{alcala14}, which span a larger range in mass. In our sample of very active cTTS, the H$\alpha$ and H$\beta$ emissions are not  good indicators because of the large optical depth and strong P Cygni profiles, which remove the blue emission wings of those lines (as seen in Figure \ref{fig1}). Therefore, we assume that the \ion{He}{I} 5876 \AA\, line is more reliable to estimate the accretion luminosity.

We first determine the ratio $L_{line}/L_{star}$ through the equation 

\begin{equation}
  \frac{L_{line}}{L_{star}} = EW * (1+VC) * \frac{F_{cont}}{L_{star,}}
\end{equation}

where $L_{star}$ is the stellar bolometric luminosity, $F_{cont}$ is the stellar continuum flux density at the wavelength of the He I line, EW is the equivalent width of the \ion{He}{I} emission, and VC is the veiling continuum near the \ion{He}{I} line.

The ratio $F_{cont}/L_{star}$ is obtained using the adequate template spectra selected from the stellar spectral flux library by \citet{pickles98}. The library covers the spectral range 1150 - 25000 \AA, and a blackbody at the star temperature is added to simulate the continuum level at wavelengths longer than 2.5 $\mu m$ for a better determination of $L_{star}$.

The ratio $L_{line}/L_{star}$ is converted to $L_{acc}/L_{star}$ using the relationship obtained by \citet{alcala14}  and the mass accretion rate determined from the equation 

\begin{equation}
 \dot{M}_{acc}= \frac{L_{acc} R_*}{G M_*}(1-R_*/R_{in})^{-1},
\end{equation}

where $R_*$ and $M_*$ are the star radius and mass, respectively, reported in Table \ref{table1}, and $R_{in}$ is the inner disk radius taken as 3$R_*$ in our calculation. Table 4 displays the final results, including the veiling continuum estimated from the plots EW/EW versus wavelength, the \ion{He}{I} line, and accretion luminosities used for the mass accretion rate determination. 

We have also checked that the accretion luminosities and the veiling continuum reported in Table \ref{table:accretion} are consistent with the hot spot hypothesis. Within the used spectral range (4350 - 6750 \AA), we do not see dependence of the veiling continuum on wavelength, which makes the estimation of the hot spot(s) temperature more difficult. We may only assert that temperatures higher than 10000 K for the spots are not expected, otherwise the slope in the plots EW/EW versus wavelength would be more pronounced. 

If we consider the typical spot temperatures for cTTS in the range 6000 - 8000 K \citep{calvet98} and assuming blackbody emission from the spot, we may infer the spot filling factor in our stars by using the equation:

\begin{equation}
 \frac{L_{acc}}{L_{star}} \sim f \left (\frac{T_{spot}}{T_{star}}  \right )^{4},
\label{eq3}
\end{equation}

where $T_{star}$ is the effective temperature of the star. 

We get filling factor values between 1 and 3 $\%$ for V1331 Cyg 7, 20 $\%$ for AS 353A, and below 1 $\%$ for Lk\ha 321. The filling factor for AS 353A seems to be unusually large, which may be caused by an overestimated veiling continuum for this star.

\begin{table*}
\caption{Determination of accretion luminosities and mass accretion rates.} % title of Table
\label{table:accretion} % is used to refer this table in the text
\centering % used for centering table
\begin{tabular}{|c|cccccc|} % centered columns (2 columns)
\hline\hline % inserts double horizontal lines
Star & Template & VC &log($L_{He}/L_{star}$) & $\log(L_{acc}/L_{star})$ & $\log(\dot{M}_{acc})$ \\ % table heading 
  &    &  & (dex) & (dex) & $(M_\odot yr^{-1})$  \\
\hline % inserts single horizontal line
Lk\ha 321 &  G8V  &  0.0 &  -4.97  & -1.72 &  -7.4 \\
V1331 Cyg  & K0V  &  0.1  &  -4.51 & -1.21 & -6.9  \\
AS 353A    & K0V  &   0.6   & -3.6 &  -0.31  &  -7.1 \\
\hline %inserts single line
\end{tabular}
\tablefoot{VC is the veiling continuum at the wavelength of the \ion{He}{I}
 line.}
\end{table*}

\subsection{Accretion-powered chromosphere}

The dependence of veiling on the strength of transition indicates that the origin of this veiling is related to a drastic change in the stellar atmosphere structure due to accretion.
The upper layers of the stellar atmosphere are most affected by the hard  radiation of the shock wave located above. Therefore, at a given accreting mass flux the cores of strong lines will be mostly affected, while the weakest lines formed in the deeper layers remain still unchanged.  In this situation, only the weakest lines may be used as a measure of the veiling continuum. 

\citet{dodin12} considered a model of veiling by lines, where the narrow emission lines, radiated in the post-shock region, contribute to the veiling along with the continuum radiation from the hot spot. The model showed that the veiling by lines prevails in cTTS with a moderate accretion flux, while with increasing accretion flux the veiling continuum becomes the most important. Our results do not contradict this prediction: the relative contribution of the veiling continuum is greater in AS 353A, where $L_{acc}/L_{star}$ is the largest of the three stars.

If the line-dependent veiling is caused by the emission lines radiated in a small post-shock region, which is hotter than the surrounding undisturbed photosphere, one would expect an emission core in the strongest lines and/or a dependence of veiling on the excitation potential. None of our objects reveals the dependence of veiling on excitation potential, probably because of the large scatter of points on the diagrams.

In AS 353A and V1331 Cyg, the \vsini is too small to resolve the details in a photospheric line profile. However, in the case of Lk\ha 321, the \vsini is larger and we can conclude from the absence of a narrow emission core in relatively broad photospheric lines that the line emission is not concentrated in a single spot but spread over a larger area above the visible stellar surface. A time series of spectral observations of this star would reveal probable variations in the photospheric line profiles. The mapping of magnetic field, cool spots, and emission line areas on the stellar surface was done for several cTTS (see \citet{donati13, johnstone14}, and references therein). In all cases there is an axial asymmetry in the distribution of the emission line areas. 

In Lk\ha 321 the line width is large enough to apply Doppler imaging.  In V1331 Cyg and AS 353A, where the photospheric lines are so narrow, it is still possible to look for variations in radial velocity caused by a passage of the presumable hot spot(s) across the visible stellar surface. So far we can only assert that there is no change in the radial velocity of V1331 Cyg within $\pm$ 0.3 \kms as derived from the HIRES spectra of 2004 and 2007.
A significant change in radial velocity was noticed only in AS 353A: Vrad = -8.0 $\pm$ 0.2 \kms in 1998 and  -10.4 $\pm$ 0.2 \kms in 2004. 

The most intriguing is that of Lk\ha 321, where the veiling continuum is absent, but the line-dependent veiling is large. A similar effect was observed in DR Tau (Petrov et al 2011). In that star both the veiling continuum and the line-dependent veiling  vary from night to night. When the veiling continuum disappears, the line-dependent veiling remains, although at a reduced level. It was interpreted as an accretion-powered  chromosphere. At a given infall velocity, the  accretion energy flux is proportional to the density of infalling gas.  At lower density the photosphere  is not  affected, while the upper layers of atmosphere are heated, thus giving rise to chromospheric-like emissions filling in the photospheric absorptions. 

Yet one more consideration concerns the magnetic field structure. The Zeeman Doppler Imaging technique enables us to restore the large-scale topology of the stellar magnetic field in cTTS (e.g. \citet{johnstone14}). So far, the Sun is the only star for which the small-scale (down to megameters) structures of magnetic fields have been investigated. It was found that the solar magnetic field has a fractal (scale-invariant) structure \citep{abramenko08}.
In fact, we expect that the local magnetic fields in cTTS within any small fraction of stellar surface may be as complex as in the Sun. In the framework of the magnetospheric accretion model, the shock front is supposed to be a confined area over a small fraction of the stellar surface. One can imagine that as the ionized infalling gas approaches the star, it meets more complicated  magnetic structures and finally enters the stellar atmosphere at the dividing lines between the local closed fields. How would this alter the shock structure and the heating of the underlying photosphere?  These small-scale processes may be important in the manifestation of the observed consequences of accretion.

%______________________________________________________________

\section{Conclusions}

The analysis of spectra of three very active cTTS of low to intermediate mass show that veiling consists of two components: a line-dependent veiling and the veiling continuum. Those two components are visible when the ratios of equivalent widths of nearby lines between template and TTS are plotted against the line strength on the template. Stronger lines are more veiled than weaker lines, regardless of wavelength and excitation potential. The veiling continuum was determined from the ratio of the  weakest lines and the accretion luminosities from the He I line emission. In the three studied stars, we found that the photospheric line veiling is dominated by the line-dependent component, while the veiling continuum is a minor contributor.
We conclude that the observed veiling on the photospheric lines of Lk\ha 321, AS 353A, and V1331 Cyg is composed of a line-dependent veiling with its origin in an abnormal structure of stellar atmosphere being heated up by the accreting matter and a veiling continuum radiated by a hot spot with temperature lower than 10000 K.

%_______________________________________________________________

\begin{acknowledgements}
      This work was supported by Fundação para a Ciência e a Tecnologia (FCT) through national funds (UID/FIS/04434/2013) and by FEDER through COMPETE2020 (POCI-01-0145-FEDER-007672). ACSR acknowledges the support of an IA fellowship: CIAAUP-17/2015-BI in the context of the project (UID/FIS/04434/2013$\&$POCI-01-0145-FEDER-007672). 
 PPP acknowledges the visitor's programme grant from Instituto de Astrofisica e Ciências do Espaço, Portugal.
 This work has made use of the VALD database, operated at Uppsala University, the Institute of Astronomy RAS in Moscow, and the University of Vienna. 
\end{acknowledgements}

%-------------------------------------------------------------------

%\begin{appendix}

%\end{appendix}


\begin{thebibliography}{}

\bibitem[Abramenko(2008)]{abramenko08} Abramenko, V.~I.\ 2008, Solar Physics Research Trends, Edited by Pingzhi Wang.~ISBN 1-60021-987-x.~ Nova Publishers, 2008, p.~95-136, 95 

\bibitem[Alcal{\'a} et al.(2014)]{alcala14} Alcal{\'a}, J.~M., Natta, A., Manara, C.~F., et al.\ 2014, \aap, 561, A2

\bibitem[Alencar \& Basri(2000)]{alencar00} Alencar, S.~H.~P., \& Basri, G.\ 2000, \aj, 119, 1881

\bibitem[Basri \& Batalha(1990)]{basri90} Basri, G., \& Batalha, C.\ 1990, \apj, 363, 654

\bibitem[Beristain et al.(1998)]{beristain98} Beristain, G.,Edwards, S., Kwan, J.\ 1998, \apj, 499, 828

\bibitem[Blanco-Cuaresma et al.(2014)]{blanco14} Blanco-Cuaresma, S., Soubiran, C., Jofr{\'e}, P., \& Heiter, U.\ 2014, \aap, 566, A98 
 
\bibitem[Calvet \& Gullbring(1998)]{calvet98} Calvet, N., \& Gullbring, E.\ 1998, \apj, 509, 802

\bibitem[Calvet et al.(2004)]{calvet04} Calvet, N., Muzerolle, J., Brice{\~n}o, C., et al.\ 2004, \aj, 128, 1294 

\bibitem[Chavarria(1981)]{chavarria81} Chavarria, C.\ 1981, \aap, 101, 105 

\bibitem[Cohen \& Kuhi(1979)]{cohen79} Cohen,  M. \&  Kuh i,  L.V. \ 1979, \apj, 41, 743                            

\bibitem[Dahm(2008)]{dahm08} Dahm, S.~E.\ 2008, \aj, 136, 521-547 

\bibitem[Dodin \& Lamzin(2012)]{dodin12} Dodin, A.~V., \& Lamzin, S.~A.\ 2012, Astronomy Letters, 38, 649

\bibitem[Donati et al.(2013)]{donati13} Donati, J.-F., Gregory, S.~G., Alencar, S.~H.~P., et al.\ 2013, \mnras, 436, 881

\bibitem[Choudhary et al.(2016)]{choudhary16} Choudhary, A., Stecklum, B., \& Linz, H.\ 2016, \aap, 590, A106

\bibitem[Eisloeffel et al.(1990)]{eisloffel90} Eisloeffel, J., Solf, J., \& Boehm, K.~H.\ 1990, \aap, 237, 369 

\bibitem[Fang et al.(2009)]{fang09} Fang, M., van Boekel, R., Wang, W., et al.\ 2009, \aap, 504, 461 

\bibitem[Fernandez \& Eiroa(1996)]{fernandez96} Fernandez, M., \& Eiroa, C.\ 1996, \aap, 310, 143 

\bibitem[Gameiro et al.(2006)]{gameiro06} Gameiro, J.~F., Folha, D.~F.~M., \& Petrov, P.~P.\ 2006, \aap, 445, 323  

\bibitem[Gahm et al.(2008)]{gahm08} Gahm, G.~F., Walter, F.~M., Stempels, H.~C., Petrov, P.~P., \& Herczeg, G.~J.\ 2008, \aap, 482, L35  

\bibitem[Grankin et al.(2007)]{grankin07} Grankin, K.~N., Melnikov, S.~Y., Bouvier, J., Herbst, W., \& Shevchenko, V.~S.\ 2007, \aap, 461, 183 

\bibitem[Hamann \& Persson(1992)]{hamann92} Hamann, F., \& Persson, S.~E.\ 1992, \apj, 394, 628

\bibitem[Hartigan et al.(1991)]{hartigan91} Hartigan, P., Kenyon, S.~J., Hartmann, L., et al.\ 1991, \apj, 382, 617

\bibitem[Hartigan et al.(1995)]{hartigan95} Hartigan, P., Edwards, S., \& Ghandour, L.\ 1995, \apj, 452, 736 

\bibitem[Herbig \& Jones(1983)]{herbig83} Herbig, G.~H., \& Jones, B.~F.\ 1983, \aj, 88, 1040

\bibitem[Herczeg \& Hillenbrand(2008)]{herczeg08} Herczeg, G.~J., \& Hillenbrand, L.~A.\ 2008, \apj, 681, 594-625

\bibitem[James et al.(2006)]{james06} James, D.~J., Melo, C., Santos, N.~C., \& Bouvier, J.\ 2006, \aap, 446, 971 

\bibitem[Johnstone et al.(2014)]{johnstone14} Johnstone, C.~P., Jardine, M., Gregory, S.~G., Donati, J.-F., \& Hussain, G.\ 2014, \mnras, 437, 3202 

\bibitem[Joy(1949)]{joy49} Joy, A.~H.\ 1949, \apj, 110, 424

\bibitem[Kolotilov(1983)]{kolotilov83} Kolotilov, E. A. \ 1983, Soviet Astronomy Letters, V. 9, P. 289

\bibitem[Kuhi(1964)]{kuhi64} Kuhi, L.V.  \apj, 140, 1409

\bibitem[Mendigut{\'{\i}}a et al.(2015)]{mendigutia15} Mendigut{\'{\i}}a, I., Oudmaijer, R.~D., Rigliaco, E., et al.\ 2015, \mnras, 452, 2837  

\bibitem[Mundt(1984)]{mundt84} Mundt, R.\ 1984, \apj, 280, 749 

\bibitem[Mundt \& Eisl{\"o}ffel(1998)]{mundt98} Mundt, R., \& Eisl{\"o}ffel, J.\ 1998, \aj, 116, 860 

\bibitem[Muzerolle et al.(1998)]{muzerolle98} Muzerolle, J., Calvet, N., \& Hartmann, L.\ 1998, \apj, 492, 743

\bibitem[Padgett(1996)]{padgett96} Padgett, D.~L.\ 1996, \apj, 471, 847

\bibitem[Petrov et al.(2001)]{petrov01} Petrov, P.~P., Gahm, G.~F., Gameiro, J.~F., et al.\ 2001, \aap, 369, 993   

\bibitem[Petrov et al.(2011)]{petrov11} Petrov, P.~P., Gahm, G.~F., Stempels, H.~C., Walter, F.~M., \& Artemenko, S.~A.\ 2011, \aap, 535, A6 

\bibitem[Petrov \& Babina(2014)]{petrovbabina14} Petrov, P.~P., \& Babina, E.~V.\ 2014, Bulletin of the Crimean Astrophysical Observatory, 110, 1

\bibitem[Petrov et al.(2014)]{petrov14} Petrov, P.~P., Kurosawa, R., Romanova, M.~M., et al.\ 2014, \mnras, 442, 3643 

\bibitem[Pickles(1998)]{pickles98} Pickles, A.~J.\ 1998, \pasp, 110, 863

\bibitem[Prato et al.(2003)]{prato03} Prato, L., Greene, T.~P., \& Simon, M.\ 2003, \apj, 584, 853

\bibitem[Quanz et al.(2007)]{quanz07} Quanz, S.P., Apai, D., Henning, Th.\ 2007, \apj, 656, 287

\bibitem[Rice et al.(2006)]{rice06} Rice, E.~L., Prato, L., \& McLean, I.~S.\ 2006, \apj, 647, 432 

\bibitem[Rigliaco et al.(2012)]{rigliaco12} Rigliaco, E., Natta, A., Testi, L., et al.\ 2012, \aap, 548, A56

\bibitem[Rojas et al.(2008)]{rojas08} Rojas, G., Gregorio-Hetem, J., \& Hetem, A.\ 2008, \mnras, 387, 1335

\bibitem[Ryabchikova et al.(2015)]{ryabchikova15} Ryabchikova, T., Piskunov, N., Kurucz, R.~L., et al.\ 2015, \physscr, 90, 054005

\bibitem[Shevchenko et al.(1991)]{shevchenko91} Shevchenko, V.~S., Yakulov, S.~D., Hambarian, V.~V., \& Garibjanian, A.~T.\ 1991, \azh, 68, 275

\bibitem[Sicilia-Aguilar et al.(2015)]{sicilia15} Sicilia-Aguilar, A., Fang, M., Roccatagliata, V., et al.\ 2015, \aap, 580, A82 

\bibitem[Siess et al.(2000)] {siess00} Siess,L., Dufour, E., \& Forestini, M. \ 2000, \aa, 358, 593

\bibitem[Taguchi et al.(2009)]{taguchi09} Taguchi, Y., Itoh, Y., \& Mukai, T.\ 2009, \pasj, 61, 251 

\bibitem[Tokunaga et al.(2004)]{tokunaga04} Tokunaga, A.~T., Reipurth, B., G{\"a}ssler, W., et al.\ 2004, \aj, 127, 444

\bibitem[Valenti et al.(1993)]{valenti93} Valenti, J.~A., Basri, G., \& Johns, C.~M.\ 1993, \aj, 106, 2024

\bibitem[Valenti $\&$ Piskunov(1996)]{valenti96}Valenti, J. A. $\&$ Piskunov, N. 1996, A$\&$AS, 118, 595

\bibitem[Valenti $\&$ Fischer(2005)]{valenti05}Valenti, J. A. $\&$ Fischer, D. A. 2005,  ApJS, 159, 141

\bibitem[Vogt et al.(1994)]{vogt94} Vogt, S.~S., Allen, S.~L., Bigelow, B.~C., et al.\ 1994, \procspie, 2198, 362

\bibitem[Welin(1976)]{welin76} Welin, G.\ 1976, \aap, 49, 145  

\end{thebibliography}
\end{document}